# Revealing electronic state-switching at conical intersections in alkyl iodides by ultrafast XUV transient absorption spectroscopy

Kristina F. Chang[1], Maurizio Reduzzi[1], Han Wang[2], Sonia M. Poullain[1,3], Yuki Kobayashi[1], Lou Barreau[1], David Prendergast[2,4], Daniel M. Neumark[1,2], and Stephen R. Leone[1,2,5,*]

[1]Department of Chemistry, University of California, Berkeley, California 94720, USA.
[2]Chemical Sciences Division, Lawrence Berkeley National Laboratory, Berkeley, California 94720, USA.
[3]Departamento de Química Física, Facultad de Ciencias Químicas, Universidad Complutense de Madrid, Madrid 28040, Spain.
[4]Molecular Foundry, Lawrence Berkeley National Laboratory, Berkeley, California 94720, USA.
[5]Department of Physics, University of California, Berkeley, California 94720, USA.
*srl@berkeley.edu

## ABSTRACT

Conical intersections between electronic states often dictate the chemistry of photoexcited molecules. Recently developed sources of ultrashort extreme ultraviolet (XUV) pulses tuned to element-specific transitions in molecules allow for the unambiguous detection of electronic state-switching at a conical intersection. Here, the fragmentation of photoexcited *iso*-propyl iodide and *tert*-butyl iodide molecules (*i*-$C_3H_7I$ and *t*-$C_4H_9I$) through a conical intersection between $^3Q_0/^1Q_1$ spin-orbit states is revealed by ultrafast XUV transient absorption measuring iodine 4*d* core-to-valence transitions. The electronic state-sensitivity of the technique allows for a complete mapping of molecular dissociation from photoexcitation to photoproducts. In both molecules, the sub-100 fs transfer of a photoexcited wave packet from the $^3Q_0$ state into the $^1Q_1$ state at the conical intersection is captured. The results show how differences in the electronic state-switching of the wave packet in *i*-$C_3H_7I$ and *t*-$C_4H_9I$ directly lead to differences in the photoproduct branching ratio of the two systems.

## Introduction

The coupled evolution of electronic and nuclear structures plays a fundamental role in molecular reactions. The dynamics of chemical reactions are traditionally viewed within the Born-Oppenheimer approximation, which assumes separation of nuclear and electronic degrees of freedom in the system. While often applicable to reactions in ground electronic states, this approximation is frequently insufficient to describe excited state dynamics following photoexcitation where extensive couplings between electron and nuclear motions arise[1–4]. At crossings between the potential energy surfaces of electronic states where conical intersections are formed, the presence of strong couplings allow a molecule to abruptly transfer from one surface to another, thereby switching its electronic character. Due to the ubiquitous occurrence of crossings among electronically excited states, the chemical outcomes of many photoinduced processes such as DNA photoprotection[5,6] and retinal isomerization in vision[7–9] are dictated by nonadiabatic state-switching at conical intersections.

Dynamics at conical intersections are often challenging to capture experimentally as they necessarily involve multiple electronic states and typically evolve on a sub-picosecond timescale. Femtosecond and attosecond transient absorption spectroscopies in the extreme ultraviolet (XUV) or soft X-ray regime, which measure resonant transitions from atomic core orbitals into unoccupied valence orbitals, offer a powerful means of resolving multi-state dynamics with excellent temporal resolution[10–16]. The sensitivity of the core-to-valence absorption features to the symmetries, orbital occupations, and spin characteristics of electronic states allows for the clear detection of state-switching associated with passage through a conical intersection. In this report, ultrafast XUV transient absorption spectroscopy measuring transitions from iodine I(4*d*) core orbitals is applied to the investigation of conical intersection dynamics in the *A*-band fragmentation of alkyl iodide molecules (R-I, R = $C_nH_m$).

The alkyl iodides constitute a valuable class of chemical systems for the investigation of nonadiabatic dynamics, as their dissociation in the *A*-band is intrinsically mediated by a conical intersection[17–24]. The *A*-band is comprised of dissociative spin-orbit states accessed by $5p \rightarrow \sigma^*$ valence excitation in the ultraviolet (UV) from a nonbonding iodine orbital into an antibonding orbital along the C-I bond. UV excitation results in rapid cleavage of the C-I bond within 200 fs[25]. Within the excitation, spin-orbit states carrying the Mulliken labels $^3Q_0$, $^1Q_1$, and $^3Q_1$ are optically accessible[17] (Supplementary Note 1). For few-carbon containing alkyl iodides, excitation to $^3Q_0$ comprises 70-80% of the oscillator strength in the *A*-band[21,26,27]. As shown schematically in **Fig. 1a**, UV excitation prepares an electronic-nuclear wave packet on the $^3Q_0$ surface correlating to the production of spin-orbit excited I*($^2P_{1/2}$) atoms. Throughout its motion along the steeply repulsive potential, a fraction of the initially prepared wave packet can cross to the $^1Q_1$ surface via a conical intersection, allowing for the release of ground state I($^2P_{3/2}$) atoms. Consequently, the production of atomic I photoproducts has been primarily attributed to nonadiabatic $^3Q_0/^1Q_1$ state-switching via the conical intersection[19,21,28–30].

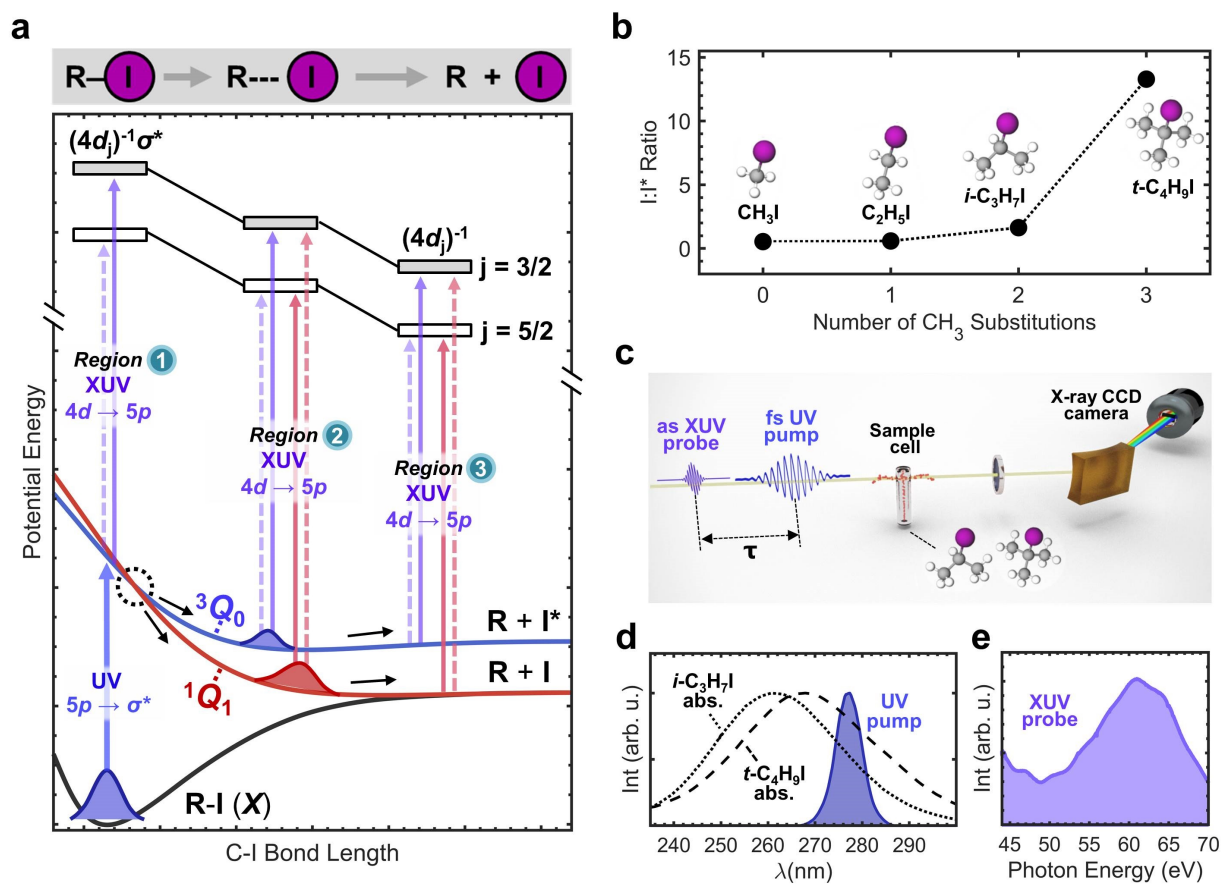

**Fig. 1.** *A*-band fragmentation of alkyl iodides and experimental outline. **a** Potential energy curves adapted from Ref.[25] are plotted as a function of C-I distance. The $^3Q_0/^1Q_1$ conical intersection (dotted circle) allows for wave packet bifurcation into I* and I dissociation channels. Dynamics along the $^3Q_0$ and $^1Q_1$ potentials before the conical intersection (*Region 1*), after the conical intersection (*Region 2*), and in the dissociation limit (*Region 3*) are mapped through XUV transitions to core-excited molecular states labeled $(4d)^{-1}\sigma^*$. The molecular core-excited states connect to atomic $(4d)^{-1}$ core-excited states at large C-I distances, and are plotted as repulsive based on their antibonding $\sigma^*$ character. **b** I:I* branching ratio data (277-280 nm excitation) obtained from Refs.[29,31–33] and plotted as a function of methyl substitutions. **c** Experimental setup, **d** UV pump spectrum and *A*-band absorption spectra of gaseous $i$-$C_3H_7I$ and $t$-$C_4H_9I$ adapted from Ref.[34], and **e** XUV probe spectrum.

The I:I* photoproduct branching ratio varies widely among alkyl iodides depending on R-group structure. Previously-measured I:I* branching ratios obtained from the dissociation of several alkyl iodides at 277-280 nm are plotted in **Fig. 1b**. Methyl and ethyl iodide ($CH_3I$ and $C_2H_5I$) dissociation forms atomic I in a minority ratio of ∼ 1:3 relative to I*[31,32]. In contrast, the dissociation of molecules with greater methyl substitution at the central carbon favor the release of atomic I. For $i$-$C_3H_7I$, atomic I photoproducts dominate in a ratio of ∼ 2:1[33], while $t$-$C_4H_9I$ provides an even greater yield of ∼ 13:1[29]. The dramatic increase in atomic I production suggests that a significantly larger fraction of the initial wave packet switches to the $^1Q_1$ surface while passing through the conical intersection[19,29,35]. While $i$-$C_3H_7I$ appears to represent an intermediate case in which the wave packet bifurcates between the $^1Q_1$ and $^3Q_0$ states in a $\sim$ 2:1 ratio, $t$-C$_4$H$_9$I appears to represent a case of nearly complete transfer to the $^1Q_1$ state. Owing to the importance of alkyl iodides as a benchmark system, direct observation of the conical intersection gating the formation of photoproducts is a very appealing target for both experiment and theory[36–38].

Here, we present the real-time mapping of wave packet bifurcation at the $^3Q_0/^1Q_1$ conical intersection in $i$-C$_3$H$_7$I and $t$-C$_4$H$_9$I using ultrafast XUV transient absorption. Experimentally, dynamics are launched by a resonant femtosecond UV pump pulse and followed by a time-delayed attosecond XUV pulse that probes transitions from I(4$d$) core orbitals (**Fig. 1a**). In the corresponding XUV absorption spectra, regions of the excited state surfaces both prior and subsequent to the conical intersection are mapped to distinct spectral features, which allows for the unambiguous detection of electronic state-switching at the $^3Q_0/^1Q_1$ conical intersection correlating to the release of atomic I* and I. The signatures of conical intersection dynamics and molecular fragmentation are found to be in excellent agreement with simulated XUV spectra of a CH$_3$I model system.

## Results

### Time-resolved probing of *iso*-propyl and *tert*-butyl iodide photodissociation

The experimental pump-probe setup is summarized in **Fig. 1c-e**. Additional details of the experimental apparatus can be found in the "Methods" section. Briefly, gaseous $i$-C$_3$H$_7$I and $t$-C$_4$H$_9$I molecules in a quasi-static gas cell are excited by UV pump pulses (277 nm, 50 fs, 5 $\mu$J per pulse) at a peak intensity of $1.1 \times 10^{12}$ W cm$^{-2}$. The UV pump spectrum is centered near the 260 nm and 268 nm *A*-band absorption maxima of $i$-C$_3$H$_7$I and $t$-C$_4$H$_9$I, respectively[34]. Following UV excitation, dynamics are probed by time-delayed isolated attosecond XUV pulses (40-70 eV, $\sim$ 170 as)[39] tuned to absorption transitions from the I(4$d$) core orbital appearing in the 45 to 48 eV photon energy range. A Gaussian instrument response function of 50 $\pm$ 7 fs (full width at half maximum) of the transient absorption experiment is measured using an *in situ* UV-XUV cross-correlation method.

As shown in **Fig. 1a**, wave packet motion from the highly repulsive region of the excited state surfaces (*Regions 1-2*) into the asymptotic dissociation limit (*Region 3*) is probed through XUV absorption transitions corresponding to core-to-valence excitations primarily localized on iodine. The $^3Q_0$ and $^1Q_1$ excited states are characterized by the configuration $(4d)^{10}\ldots(\sigma)^2(5p)^3(\sigma^*)^1$ where the nonbonding 5$p$ valence orbitals on iodine possesses 5$p\pi^*$ character due to interactions with the alkyl moiety. In a one-electron transition picture[36,40,41], the excited states can be probed by the excitation of available 4$d\rightarrow$5$p$ transitions to distinct $(4d_{3/2})^{-1}\sigma^*$ and $(4d_{5/2})^{-1}\sigma^*$ core-excited states separated in energy by the 4$d$ core-hole spin-orbit splitting. In the corresponding XUV absorption spectra, transitions appear as doublets with excitations to $(4d_{3/2})^{-1}\sigma^*$ appearing at higher photon energies compared to excitations to $(4d_{5/2})^{-1}\sigma^*$. In **Fig. 1a**, stronger and weaker XUV transitions are distinguished by solid and dashed arrows. In this study, the $^3Q_0$ state is observed to primarily undergo strong transitions to the $(4d_{3/2})^{-1}\sigma^*$ state appearing at higher XUV energies in the spectrum, whereas the $^1Q_1$ state primarily undergoes strong transitions to the $(4d_{5/2})^{-1}\sigma^*$ state appearing at lower XUV energies. In addition to their primary appearance at distinct photon energies, dynamics along the $^3Q_0$ and $^1Q_1$ potentials are furthermore distinguished by their evolution at long time delays. During molecular fragmentation along the C-I bond, the collapse of the hybridized molecular orbitals surrounding the iodine atom leading to a purely-atomic $(4d)^{10}\ldots(5p)^5$ configuration is spectroscopically revealed through the convergence of molecular $^3Q_0$ and $^1Q_1$ features into peaks associated with free I* and I atoms at long time delays.

Transient absorption spectra are recorded as changes in optical density $\Delta$OD = $-\log[I_{XUV+UV}(E,\tau)/I_{XUV}(E)]$, where $I_{XUV+UV}(E,\tau)$ is the XUV spectrum recorded at the time delay $\tau$ following the UV pump and $I_{XUV}(E)$ is the XUV spectrum recorded in the absence of the pump. The scan averages and integration times used to record the experimental transients are described in the "Methods" section. To eliminate high-frequency noise, the recorded transients are post-processed using a low pass filter (Supplementary Note 2). After post-processing, the experimental noise level is estimated as $\sim$ 2 mOD. In **Fig. 2a-d**, the resulting transients for $i$-C$_3$H$_7$I and $t$-C$_4$H$_9$I are plotted between 44.5 and 48.5 eV photon energies where time-dependent features that reflect excited state dynamics are observed. Spectra plotted over the full photon energy range recorded (44 to 60 eV) can be found in Supplementary Fig. 1-2.

Experimental spectra plotted at time delay intervals between -4 to 160 fs are shown in **Fig. 2a-b**. Several discrete, time-dependent features are observed in the spectra. The rich evolution of the features can be observed in the colormap depictions of the transient spectra shown in **Fig. 2c-d**. The convergence of features at early time delays (0 to 100 fs) into the fixed values of atomic transitions at longer times (100 to 160 fs) reflects dynamics evolving from the steeply repulsive to the asymptotic regions of the excited state potentials. In particular, dissociation in the asymptotic region (*Region 3*, **Fig. 1a**) is signified by the rise of well-known atomic transitions at 45.9 eV [I($^2P_{3/2} \rightarrow\ ^2D_{5/2}$)], 46.7 eV [I*($^2P_{1/2} \rightarrow\ ^2D_{3/2}$)], and 47.6 eV [I($^2P_{3/2} \rightarrow\ ^2D_{3/2}$)][42,43]. In the $i$-C$_3$H$_7$I transient, atomic I and I* peaks are clearly visible whereas in the $t$-C$_4$H$_9$I transient, only the atomic I peaks are observed. The intensities of the observed atomic transitions allow I:I* branching ratio estimates of $\sim$ 2:1 for $i$-C$_3$H$_7$I and $\geq$ 9:1 for $t$-C$_4$H$_9$I using the assumption that the I* signal is below the 2 mOD noise level of the experiment (Supplementary Note 3), and are consistent with the I:I* yields reported by previous measurements[29,33].

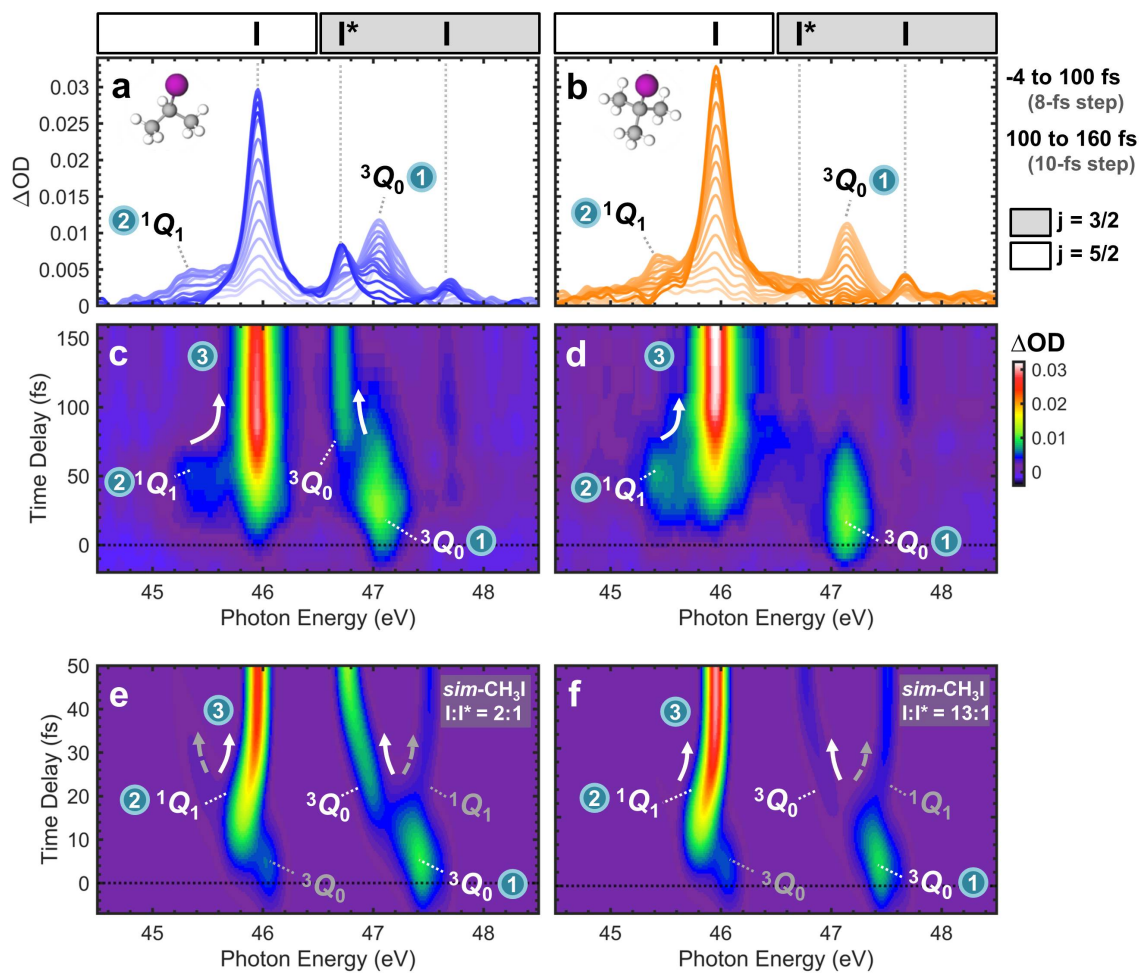

**Fig. 2. Experimental *i*-C$_3$H$_7$I and *t*-C$_4$H$_9$I transients and modified simulations of CH$_3$I transients. a,b** Experimental spectra taken at selected time delays between -4 to 160 fs for *i*-C$_3$H$_7$I (*left, blue*) and *t*-C$_4$H$_9$I (*right, orange*). Later time delays are plotted in darker colors. **c,d** Experimental transients for *i*-C$_3$H$_7$I (*left*) and *t*-C$_4$H$_9$I (*right*). **e,f** Modified transient simulations of CH$_3$I converging to the empirical I:I* branching ratios from *i*-C$_3$H$_7$I (*left*) and *t*-C$_4$H$_9$I (*right*). The simulations are temporally broadened by a Gaussian and independently normalized for comparison to the experiments. Dashed vertical lines indicate the positions of atomic iodine transitions. State-specific molecular features and their convergence (indicated by arrows) to the atomic transitions are labeled according to the *Region 1-3* labeling scheme introduced in **Fig. 1a**.

## Spectroscopic mapping of conical intersection dynamics preceding fragmentation

Molecular features located at distinct photon energies from the atomic peaks in the XUV spectrum reveal dynamics in the repulsive region of the excited state potentials where the conical intersection is found (*Regions 1-2*, **Fig. 1a**). In both molecules (**Fig. 2a-d**), discrete molecular features located at 47.1 and 45.4 eV are transiently observed. While the two features both rise and decay within 100 fs, they otherwise exhibit different dynamics. The 47.1 eV feature is observed to maximize in intensity within the instrument response function. In contrast, the 45.4 eV feature maximizes in intensity 25 fs later, and its rise thus accompanies the decay of the 47.1 eV feature. Furthermore, the two features converge to different atomic absorption lines in the long-time limit. In the *i*-C$_3$H$_7$I transient where both I* and I peaks are observed, the 47.1 and 45.4 eV features exhibit clear shifts in energy into the I*(46.7 eV) and I(45.9 eV) lines, respectively. In the *t*-C$_4$H$_9$I transient where only I peaks are observed, the 45.4 eV feature shifts into the I(45.9 eV) absorption line whereas the 47.1 eV feature disappears with no accompanying I* rise. The molecular features at 47.1 and 45.4 eV are therefore assigned to $^3Q_0$ and $^1Q_1$ states, respectively. In accordance with its prompt appearance, the 47.1 eV ($^3Q_0$) feature is assigned to the region before the conical intersection (*Region 1*) directly populated by the UV pump. Meanwhile, the 45.4 eV ($^1Q_1$) feature is ascribed to the region after the conical intersection (*Region 2*) populated through nonadiabatic transitions from the $^3Q_0$ state at later times. The determination of time constants characterizing the evolution of the $^3Q_0$ (*Region 1*) and $^1Q_1$ (*Region 2*) features is currently limited by the temporal resolution of the experiments and is therefore not provided here. However, the observed decay in intensity of the $^3Q_0$ (*Region 1*) and $^1Q_1$ (*Region 2*) features indicates that passage through the conical intersection is complete on a sub-100 fs timescale.

The concomitant decay and rise of the state-specific molecular signals in the XUV, appearing as a discontinuous switching of intensity from the $^3Q_0$ (*Region 1*) feature to the $^1Q_1$ (*Region 2* feature in the transients, provides a clear spectroscopic signature of electronic reconfiguration at a conical intersection. Following passage through the conical intersection, dissociation dynamics along the $^3Q_0$ and $^1Q_1$ potentials (*Regions 2-3*) are mapped through continuous shifts into corresponding atomic I* and I absorption lines in the long-time limit. Thus, the ability of XUV spectroscopy to temporally and energetically resolve all electronic states involved in the reaction allows for the complete mapping of molecular fragmentation dynamics in the *A*-band, including the critical moment of wave packet bifurcation at the conical intersection. The experimental XUV transients obtained exemplify the cases of intermediate and nearly complete electronic reconfiguration at a conical intersection. In the case of $i$-C$_3$H$_7$I (**Fig. 2c**), intermediate state-switching resulting in the partial retention of population on $^3Q_0$ after the conical intersection is directly observed as the spectral feature connecting the $^3Q_0$ (*Region 1*) signal to the atomic I* limit. In contrast, the nearly complete transfer of population from $^3Q_0$ to $^1Q_1$ in $t$-C$_4$H$_9$I (**Fig. 2d**) is signified by the abrupt disappearance of the $^3Q_0$ (*Region 1*) signal.

As shown in the experimental schematic in **Fig. 1a**, dynamics are followed through excitations to $(4d_{3/2})^{-1}\sigma^*$ and $(4d_{5/2})^{-1}\sigma^*$ core-excited states. The 47.1 eV ($^3Q_0$) signal can be assigned to $(4d_{3/2})^{-1}\sigma^*$ excitation based on its I*($^2P_{1/2} \to$ $^2D_{3/2}$) convergence limit. Similarly, the 45.4 eV ($^1Q_1$) feature can be assigned to $(4d_{5/2})^{-1}\sigma^*$ excitation based on its I($^2P_{3/2} \to$ $^2D_{5/2}$) convergence limit. In principle, complementary $^3Q_0$ and $^1Q_1$ transitions associated with $(4d_{5/2})^{-1}\sigma^*$ and $(4d_{3/2})^{-1}\sigma^*$ excitations, respectively, are also possible. Such transitions would appear as continuous signals connecting the 45.4 eV ($^1Q_1$) feature at early times to the atomically-forbidden 45.0 eV [I*($^2P_{1/2} \to$ $^2D_{5/2}$)] convergence limit, and connecting the 47.1 eV ($^3Q_0$) feature to the weakly-allowed 47.6 eV [I($^2P_{3/2} \to$ $^2D_{3/2}$)] convergence limit. However, no such signals are observed in the experimental results (**Fig. 2a-d**) and are presumed to be too weak to be detected. Instead, the selective probing of $^3Q_0$ transitions to the $(4d_{3/2})^{-1}\sigma^*$ core-excited state, and $^1Q_1$ transitions to the $(4d_{5/2})^{-1}\sigma^*$ core-excited state results in the resolution of state-specific features at energetically distinct locations in the spectrum, thereby giving rise to a discontinuous appearance of electronic state-switching in the XUV spectra.

The assigned signatures of conical intersection and dissociation dynamics in $i$-C$_3$H$_7$I and $t$-C$_4$H$_9$I show strong resemblances to one another, as well as to those found in previously-simulated CH$_3$I spectra. Similarities between the positions of the state-specific XUV features in the $i$-C$_3$H$_7$I and $t$-C$_4$H$_9$I spectra are consistent with expected similarities between their valence and core-excited states (Supplementary Note 4), and motivate further comparisons to computed CH$_3$I spectra. Simulated XUV transients representing the *A*-band dissociation of CH$_3$I are directly obtained from Ref.[38]. To facilitate comparisons to the experiments, the simulations are modified to reflect partial (I:I* = 2:1) and nearly-complete (I:I* = 13:1) electronic state changes upon passage through the conical intersection. The modified CH$_3$I transients (**Fig. 2e-f**) qualitatively reproduce the spectroscopic signatures observed in the experimental results with excellent agreement. As in the experiments, bond-breaking dynamics are revealed through the convergence of molecular features into atomic limits. In the CH$_3$I simulations, the convergence is completed more quickly as compared to $i$-C$_3$H$_7$I and $t$-C$_4$H$_9$I, consistent with more rapid CH$_3$I fragmentation[25]. The spectroscopic signature of $^3Q_0/^1Q_1$ conical intersection dynamics as a rise and decay of well-separated $^1Q_1$ and $^3Q_0$ features is also reproduced. As in the experiments, the selectivity of transitions from the $^3Q_0$ and $^1Q_1$ states to the $(4d_{3/2})^{-1}\sigma^*$ and $(4d_{5/2})^{-1}\sigma^*$ core-excited states, respectively, results in a characteristically discontinuous appearance of $^3Q_0/^1Q_1$ state-switching in the XUV spectrum.

## Discussion

Conical intersection dynamics in the alkyl iodides have long been a prototype for understanding nonadiabatic processes in photochemistry. In this work, nonadiabatic fragmentation dynamics of $i$-C$_3$H$_7$I and $t$-C$_4$H$_9$I are revealed by ultrafast XUV transient absorption spectroscopy. In both molecules, spectroscopic measurements from the perspective of core-to-valence excitations localized on iodine allow for a complete mapping of the chemical reaction from UV photoexcitation to photoproduct formation. The sensitive detection of transient molecular and atomic electronic states involved in the fragmentation pathway provides an exacting picture of ultrafast wave packet bifurcation between electronic states at a conical intersection. Specifically, XUV signatures portraying the cases of partial wave packet transfer in $i$-C$_3$H$_7$I leading to an intermediate I:I* branching ratio and nearly complete wave packet transfer in $t$-C$_4$H$_9$I leading to the dominant formation of I atoms are recovered. Furthermore, by comparisons to calculated spectra of a CH$_3$I model system, the XUV signatures are shown to be readily interpretable within a straightforward, one-electron picture of core-to-valence transitions.

The present study demonstrates the general advantages of resonant photoexcitation combined with a direct probing of valence electronic structure in the XUV for capturing nonadiabatic electronic state-switching in polyatomic systems. Future experiments with shorter pump pulses, achieving faster temporal resolution, will allow time constants characterizing nonadiabatic population transfer in $i$-$C_3H_7I$ and $t$-$C_4H_9I$ to be measured and will be the subject of a future study. In addition, complementary experiments probing core-level transitions at the carbon K-edge[10,11,44] could allow for the detection of structural dynamics within the R-group moiety of the alkyl iodides during C-I dissociation, thus providing a multidimensional picture of passage through the conical intersection. Finally, the application of ultrafast XUV transient absorption methodologies to classes of molecules beyond alkyl iodides will continue to provide a powerful route for the direct investigation of non-Born-Oppenheimer dynamics governing the chemistry of electronically-excited systems.

## Methods

### Experimental setup

The $i$-$C_3H_7I$ and $t$-$C_4H_9I$ molecules are obtained from Sigma-Aldrich at 99% and 95% purity, respectively. The sample target consists of a 3 mm long quasi-static gas cell filled to a pressure of $\sim$ 5 Torr at room temperature (298 K). Alkyl iodide molecules in the gas phase are excited by UV pump pulses and probed by time-delayed attosecond XUV pulses.

Attosecond XUV probe pulses are generated by a table-top high harmonic setup. A detailed description of the high harmonic setup can be found in Ref.[39]. Briefly, the setup employs the output of a carrier-envelope phase stable Ti:Sapphire amplifier delivering 27 fs, near-infrared (NIR) pulses at a 1 kHz repetition rate. Spectral broadening of the pulses in a hollow-core fiber filled with neon, and subsequent compression by a combination of chirped mirrors and passage through an ammonium dihydrogen phosphate crystal and fused silica produces few-cycle, sub-4 fs pulses with a broadband spectrum extending from 500 to 900 nm. By focusing the few-cycle NIR pulses into a quasi-static gas cell filled with argon, isolated attosecond XUV pulses are generated through amplitude gating. The spectrum of the attosecond XUV pulses exhibits a smooth continuum structure between 40 and 70 eV. According to previous streaking measurements, the XUV pulse duration is estimated to be $\sim$ 170 as.[39] Residual NIR light is subsequently removed from the XUV beam path by a 200 nm thick aluminum filter. The XUV pulses are then focused into the sample gas cell by a toroidal mirror, and the transmitted spectrum is spatially dispersed by a concave grating and measured by an x-ray CCD camera. In this study, the photon energy range between 44 and 60 eV is mainly employed. The photon energies of the XUV spectrum are calibrated using well-known Fano resonances of neon between 40 and 50 eV.[45] By fitting the resonances to a Fano lineshape convolved with a Gaussian function representing the experimental spectral resolution, a spectral resolution of 40 meV (full width at half maximum) is estimated.

Femtosecond UV pump pulses are generated by a sum-frequency mixing method between broadband and narrowband pulses[46] briefly outlined in Supplementary Note 5. The UV pump arm is loosely-focused to a spot size of 90 $\mu$m in the sample gas cell at a crossing angle of 0.7° with respect to the XUV arm. The UV beam after the sample cell is blocked before the x-ray CCD camera by a 200 nm thick aluminum filter. Time overlap of the UV pump and XUV probe pulses as well as the UV-XUV instrument response function are characterized *in situ* via the measurement of ponderomotive shifts in core-excited atomic xenon with the UV pulse modeled as a Gaussian[36,40]. Following this methodology, an instrument response function of 50 $\pm$ 7 fs is determined. Based on this, the UV pump pulses are anticipated to be $\sim$ 50 fs in duration at the gas target.

Each time-dependent XUV spectrum of the $i$-$C_3H_7I$ experiment (**Fig. 2a,c**) and $t$-$C_4H_9I$ experiment (**Fig. 2b,d**) is obtained from an average of 70 and 50 x-ray camera frames, respectively. Each frame is captured at an integration time of 1 second per frame, 1000 laser pulses per second. In both experiments, XUV spectra are collected at time delays from -50 to 160 fs. Between -20 to 100 fs time delays, spectra are recorded at 4-fs intervals. Outside of this delay window (i.e. -50 to -20 fs and 100 to 160 fs), spectra are recorded at 10-fs intervals. The average standard deviation in ΔOD across the 44-60 eV photon energy range of interest in the XUV spectrum is $\sigma_{avg}$ = 2 mOD and is interpreted as the noise level of the experiments.

### Simulation details

Data from the simulations used to produce theoretical $CH_3I$ transients for comparison to the experimental $i$-$C_3H_7I$ and $t$-$C_4H_9I$ results are published and fully described in Ref.[38]. Briefly, nonadiabatic dynamics of $CH_3I$ after photoexcitation to the $^3Q_0$ state were simulated with Tully's fewest switches surface hopping theory as implemented in the SHARC software package. The resulting molecular dynamics trajectory data were used for the computation of XUV transient absorption spectra simulated with OpenMolcas using the MS-CASPT2 method and ANO-RCC-VTZP basis set. Molecular trajectories leading to the dissociation of I* and I atoms provide distinct signatures in the XUV transients which are plotted in Supplementary Fig. 5. Direct sums of the XUV transients associated with I* and I dissociation were used to produce the modified $CH_3I$ transients shown in **Fig. 2e-f**.

## Acknowledgments

This work is supported by the National Science Foundation (NSF) (No. CHE-1361226 and No. CHE-1660417) (K.F.C., M.R., Y.K., S.R.L.), the U.S. Army Research Office (ARO) (No. W911NF-14-1-0383) (K.F.C., Y.K., D.M.N., S.R.L.), and the U.S. Department of Energy (DOE) (No. DE-AC02-05CH11231) (H.W., D.P.). Supercomputer time was provided by the National Energy Research Scientific Computing Center (NERSC). S.M.P. acknowledges funding from the European Union's Horizon 2020 research and innovation program under the Marie Sklodowska-Curie grant agreement (No. 842539, ATTO-CONTROL). Y.K. acknowledges financial support by the Funai Overseas Scholarship. L.B. acknowledges the support of a Fellowship from the Miller Institute for Basic Research in Science, University of California Berkeley. We also thank A. Zanchet and A.R. Attar for useful scientific discussions, as well as C. Manzoni and R. Borrego-Varillas for discussions regarding the construction of the pump pulse setup.

## Author contributions

K.F.C., D.M.N., and S.R.L. conceived the experiments. K.F.C. performed the experimental measurements and analyzed the results. M.R. and K.F.C. constructed the experimental pump pulse setup. H.W. and D.P. performed the theoretical calculations. K.F.C., D.N.M., and S.R.L. wrote the manuscript with inputs from all authors.

## Additional information

**Supplementary Information** accompanies this paper.
**Competing interests** The authors declare no competing financial interests.

## Data availability

The data supporting the findings of this study are contained in the manuscript and can be made available from the corresponding author upon reasonable request.

## Code availability

The codes used to analyze the experimental results are available from the corresponding author upon reasonable request.

Online Supplementary Material for "Revealing electronic state-switching at conical intersections in alkyl iodides by ultrafast XUV transient absorption spectroscopy"

Chang *et al.*

## Supplementary Note 1: Nonadiabatic couplings in the alkyl iodide *A*-band

The excited states of alkyl iodides in the *A*-band are conventionally labeled using Mulliken notation (i.e. $^1Q_1$, $^3Q_0$, and $^3Q_1$) intended for molecules of $C_{3v}$ symmetry[1–3]. Nonadiabatic interactions between the $^3Q_0$ and $^1Q_1$ states are of primary interest in this work. For alkyl iodides within the $C_{3v}$ symmetry group such as *t*-$C_4H_9I$, $^3Q_0$ and $^1Q_1$ represent excited states of $A_1$ and E symmetry, respectively[4] while for alkyl iodides within the $C_s$ symmetry group such as *i*-$C_3H_7I$, $^3Q_0$ transforms into an A' state and $^1Q_1$ splits into the doubly-degenerate [A', A"] states[5]. Taking into account the potential energy curves reported previously for *i*-$C_3H_7I$ and *t*-$C_4H_9I$[3], we adopt the commonly-used labels of $^3Q_0$ and $^1Q_1$ throughout the main text and refer to the region of the potential energy surfaces facilitating nonadiabatic population transfer between the states as a conical intersection.

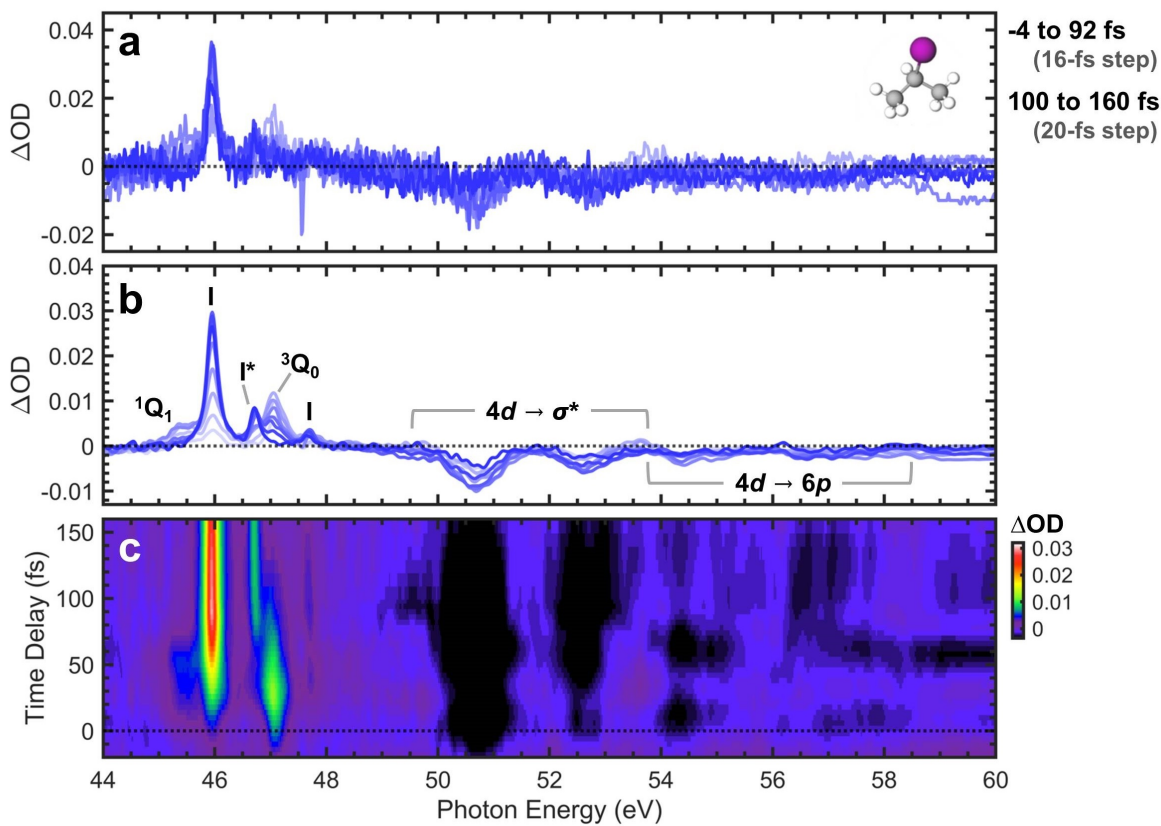

**Supplementary Fig. 1. Experimental transient absorption spectra of $i$-C$_3$H$_7$I plotted between 44 and 60 eV.**
Stacked-plot representation of **a** raw and **b** post-processed experimental transient spectra taken at selected time delays between -4 to 160 fs. Later time delays are plotted in darker colors. **c** Colormap representations of the post-processed transient spectra. Positive ΔOD features (bright shades) appearing between 44 and 49 eV are assigned to $4d \to 5p$ transitions associated with $^3Q_0$, $^1Q_1$, I*, and I species. Negative ΔOD features (dark shades) appearing between 49 and 60 eV are assigned to depleted transitions from the electronic ground state. The observed ground state features are consistent with previous absorption measurements in $i$-C$_3$H$_7$I[8], and are assigned to the series of $4d \to \sigma^*$ and $4d \to 6p$ excitations.

## Supplementary Note 2: Low pass noise filtering procedure

The raw experimental spectra are post-processed using a low pass Gaussian filter with an 8 fs and 40 meV cutoff. The procedure effectively filters out high-frequency fluctuations faster than 8 fs, which is much faster than the 50 fs Gaussian instrument response of the experiment. In addition, the procedure filters out oscillatory features in the XUV spectrum narrower than 40 meV, which are primarily attributed to noise. In the XUV spectra, atomic iodine $4d \to 5p$ transitions are expected to give rise to the narrowest features. According to previous measurements, the Lorentzian linewidths of atomic iodine $4d \to 5p$ peaks exceed 150 meV[6,7]. Molecular and atomic XUV features of interest in the present work are therefore unlikely to be accidentally removed by the 40 meV cut-off of the Gaussian filter. Raw and post-processed experimental spectra are shown in **Supplementary Fig. 1,2**. The experimental characterizations of spectral resolution and of the instrument response function discussed in the "Methods" section of the main text are performed on data sets subjected to the filtering procedure described above.

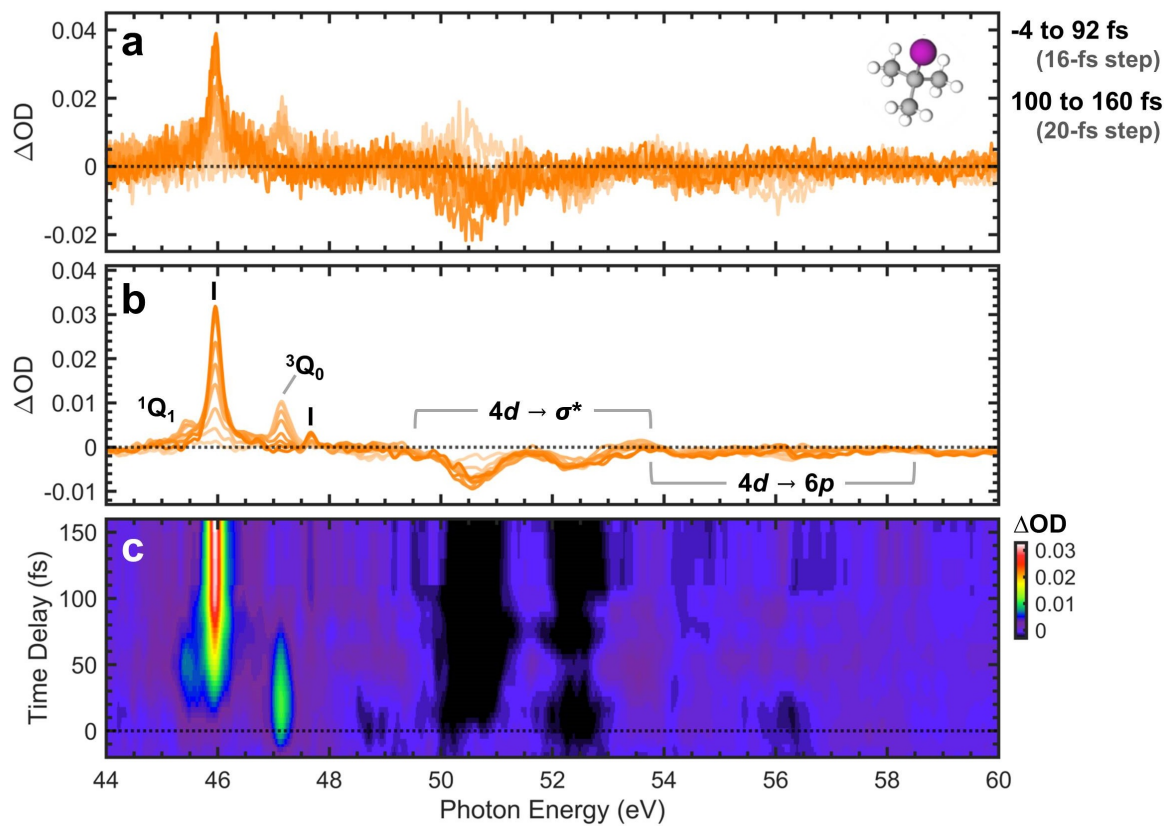

**Supplementary Fig. 2.** Experimental transient absorption spectra of *t*-$C_4H_9I$ plotted between 44 and 60 eV. Stacked-plot representation of **a** raw and **b** post-processed experimental transient spectra taken at selected time delays between -4 to 160 fs. Later time delays are plotted in darker colors. **c** Colormap representations of the post-processed transient spectra. Positive ΔOD features (bright shades) appearing between 44 and 49 eV are assigned to $4d \rightarrow 5p$ transitions associated with $^3Q_0$, $^1Q_1$, I*, and I species. Negative ΔOD features (dark shades) appearing between 49 and 60 eV are assigned to depleted transitions from the electronic ground state. The observed ground state features are qualitatively similar to those previously measured in *i*-$C_3H_7I$[8] and $CH_3I$[9] and are assigned to $4d \rightarrow \sigma^*$ and $4d \rightarrow 6p$ excitations.

## Supplementary Note 3: Estimation of I:I* photofragment branching ratio

The branching ratios between the I($^2P_{3/2}$) and I*($^2P_{1/2}$) atomic photoproducts of *i*-$C_3H_7I$ and *t*-$C_4H_9I$ fragmentation are estimated from the XUV spectra captured at 160 fs time delay. Intensities of the atomic signals in the XUV spectrum, S(I) and S(I*), are extracted as the ΔOD amplitudes of Voigt functions fit to the I(45.9 eV) and I*(46.7 eV) peaks. The I:I* photoproduct branching ratio is then calculated according to the formula:

$$I:I^* = k_{I^*/I}\, S(I) / S(I^*) \quad (Eq.\ 1)$$

where $k_{I^*/I}$ is the relative ratio of oscillator strengths associated with the I*(46.7 eV) and I(45.9 eV) transitions. A value of $k_{I^*/I} = 0.55$ is obtained from oscillator strength values reported in Supplementary Ref.[7]. Based on this procedure, the I:I* the photoproduct branching ratio is estimated to be ∼ 2:1 for *i*-$C_3H_7I$. In the case of *t*-$C_4H_9I$, the I* signal is not detected by the fitting and is presumed to be less intense than the estimated noise-level (2 mOD) of the experiment. That is, the I* signal intensity has the following upper bound: S(I*) ≤ 2 mOD. Therefore, a lower bound for the photoproduct branching ratio of t-$C_4H_9I$ can be obtained as I:I* ≥ 9:1.

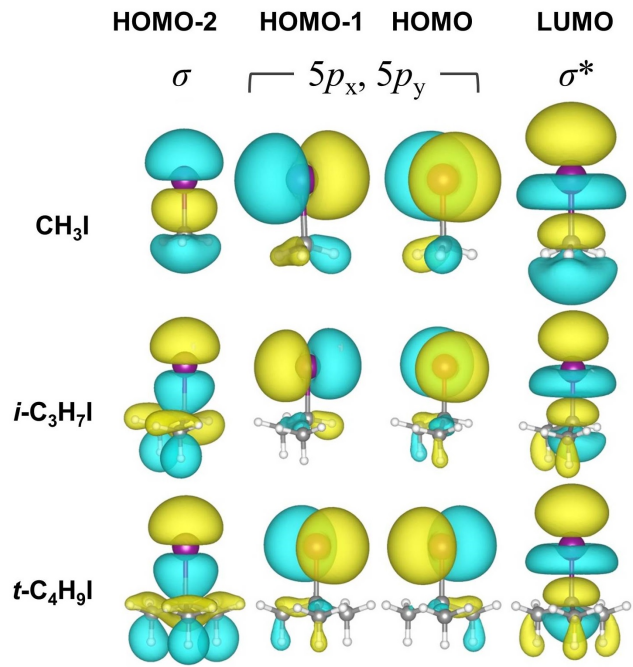

**Supplementary Fig. 3. The HOMO-2, HOMO-1, HOMO and LUMO orbitals of ground state $CH_3I$, $i$-$C_3H_7I$, and $t$-$C_4H_9I$.** In all three alkyl iodides, the HOMO-1 and HOMO orbitals correspond to $5p_x$ and $5p_y$ nonbonding orbitals localized on iodine which contain $\pi^*$ character due to interactions with the alkyl moiety, while the HOMO-2 and LUMO orbitals correspond to $\sigma$ bonding and $\sigma^*$ antibonding orbitals oriented along the C-I bond, respectively. The orbitals are calculated by DFT, using the PBE0 functional in NWChem. The iso-surface of the plotted orbitals is 0.03.

## Supplementary Note 4: Comparisons between the XUV absorption spectra of $CH_3I$, $i$-$C_3H_7I$, and $t$-$C_4H_9I$

In this work, the XUV absorption spectra of $CH_3I$, $i$-$C_3H_7I$, and $t$-$C_4H_9I$ are compared. As pictured schematically in Fig. 1a of the main text, the positions of features in the XUV spectra between 44 and 48 eV correspond to energy differences between valence states ($^3Q_0$ and $^1Q_1$) and core-excited states (($4d$)$^{-1}\sigma^*$) to which the valence states are probed. According to previous calculations, the potential energy curves of the valence $^3Q_0$ and $^1Q_1$ states of $CH_3I$, $i$-$C_3H_7I$, and $t$-$C_4H_9I$ are effectively identical along the C-I bond-breaking coordinate[3]. Calculations of the core-excited potential energy curves, corresponding to the configuration ($4d$)$^{-1}\sigma^*$, have not been performed for the molecules in this study. However, due to the effective invariance between the molecular valence orbitals (**Supplementary Fig. 3**) and the iodine I($4d$) core-orbitals of the alkyl iodides, the shapes of the core-excited potential energy curves are also expected to be similar. Energy differences between the valence and core-excited states are therefore expected to be relatively invariant between the alkyl iodides, giving rise to similar XUV spectra.

Consistent with the expectations described above, XUV features associated with the molecular $^3Q_0$ and $^1Q_1$ states appear at identical photon energies in the experimental $i$-$C_3H_7I$ and $t$-$C_4H_9I$ spectra, measured within a resolution of 40 meV. In both molecules, $^3Q_0$ (*Region 1*) and $^1Q_1$ (*Region 2*) features are observed at 47.1 and 45.4 eV, respectively. In simulated $CH_3I$ spectra based on Supplementary Ref.[12], $^3Q_0$ (*Region 1*) and $^1Q_1$ (*Region 2*) features appear at 47.5 and 45.8 eV, respectively. Molecular features in the simulations therefore appear up to 0.4 eV higher in photon energy. However, the qualitative signatures of alkyl iodide dissociation along the C-I bond and wave packet bifurcation at the $^3Q_0/^1Q_1$ conical intersection in the XUV, which is the central focus of the present study, remain in excellent agreement between the simulated $CH_3I$ spectra and measured $i$-$C_3H_7I$ and $t$-$C_4H_9I$ results.

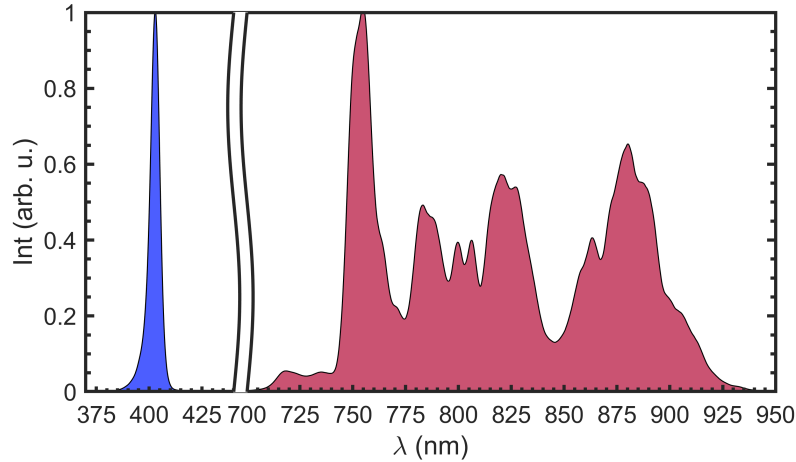

**Supplementary Fig. 4.** Spectrum of the narrowband 400 nm (shaded blue) and broadband near-infrared pulses (shaded pink) used for the sum-frequency generation of UV pump pulses.

## Supplementary Note 5: Generation of UV pump pulses

UV pump pulses are generated by sum-frequency mixing of narrowband 400 nm and broadband near-infrared (NIR) pulses[10]. The setup is initially based on the output of a carrier-envelope phase stable Ti:Sapphire amplifier delivering 27 fs, narrowband near-infrared (NIR) pulses at a 1 kHz repetition rate. A small portion of the narrowband NIR output is frequency-doubled to generate narrowband 400 nm pulses in a 200 $\mu$m thick barium borate (BBO) crystal cut at $\theta = 29.2°$ and $\phi = 90°$ for Type I phase-matching. The remaining narrowband NIR output is spectrally broadened in a hollow-core fiber and subsequently compressed in time to yield broadband, few-cycle NIR pulses with a spectrum between 500-900 nm[11]. A portion of the output is subsequently transmitted through a dichroic beamsplitter to obtain spectrally-filtered NIR pulses with a spectrum between 750-900 nm. The spectra of the obtained narrowband 400 nm and broadband NIR (750-900 nm) pulses are shown in **Supplementary Fig. 4**. Sum-frequency mixing between the pulses in a 50 $\mu$m thin BBO crystal cut at $\theta = 41.1°$ and $\phi = 90°$ for Type I phase-matching generates UV pulses. Following sum-frequency generation, residual 400 nm and NIR light is removed from the UV beam path by three reflective dielectric mirrors. The isolated UV pulses are estimated to be 50 fs in duration (see "Methods" section of the main text for additional details) with a spectrum centered at 277 nm and pulse energies of up to 7 $\mu$J per pulse.

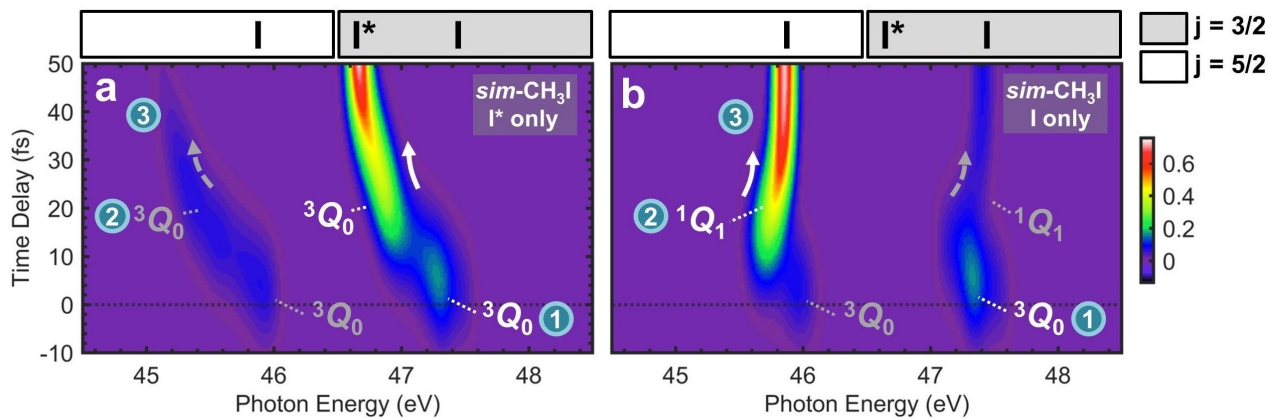

**Supplementary Fig. 5. Simulated CH$_3$I transients directly obtained from Supplementary Ref.[12]. a** Diabatic dissociation on the $^3Q_0$ state, resulting in the release of I*. **b** Adiabatic dissociation involving state-switching at the conical intersection (∼ 10-20 fs) to the $^1Q_1$ state, resulting in the release of I. To allow for comparisons with the experimental studies in the main text, the transients are temporally broadened by a Gaussian (15 fs, full width at half maximum). The colorscales of the plots are independently normalized. State-specific molecular features and their shifts (indicated by arrows) into the atomic I* and I limits are labeled according to the scheme presented in Fig. 1a of the main text.

## Supplementary Note 6: Simulated CH$_3$I transients

Simulated CH$_3$I transients directly obtained from Supplementary Ref.[12] are plotted in **Supplementary Fig. 5**. The simulations correspond to molecular dynamics trajectories initiated on the $^3Q_0$ state at 0 fs time delay, that evolve along diabatic and adiabatic reaction paths leading to the formation of I* and I, respectively. The adiabatic and diabatic dissociation paths give rise to distinct XUV transients. Direct sums between the two representative XUV transients, enforcing the convergence of I:I* branching ratios to values of 2:1 and 13:1 at 50 fs time delay, were taken in order to produce the modified transients shown in Fig. 2e-f of the main text.

## Supplementary References